\documentclass[a4paper,fleqn,usenatbib]{mnras}

\usepackage{newtxtext,newtxmath}

\usepackage[T1]{fontenc}
\usepackage{ae,aecompl}

\usepackage{times}
\usepackage{amssymb}
\usepackage{amsmath}
\usepackage{color}
\usepackage{bm}
\usepackage{graphicx}
\usepackage{tabu}
\usepackage{epsfig}
\usepackage{longtable,tabu}
\usepackage{url}
\usepackage{xcolor}
\usepackage{ulem}
\usepackage{supertabular}
\usepackage{caption}
\usepackage[multidot]{grffile}
\usepackage{subcaption}
\DeclareSymbolFont{matha}{OML}{txmi}{m}{it}
\DeclareMathSymbol{\varv}{\mathord}{matha}{118}

\def\mnras{MNRAS}

\def\aap{A\&A}
\def\apj{ApJ}
\def\apjl{ApJ}
\def\apjs{ApJS}
\def\araa{ARA\&A}

\def\nat{Nature}

\def\l1.4{$L_{\rm 1.4GHz}$} \def\s1.4{$S_{\rm 1.4GHz}$}

\def\um{$\mu$m}

\def\gs{\mathrel{\raise0.35ex\hbox{$\scriptstyle >$}\kern-0.6em
\lower0.40ex\hbox{{$\scriptstyle \sim$}}}}
\def\ls{\mathrel{\raise0.35ex\hbox{$\scriptstyle <$}\kern-0.6em
\lower0.40ex\hbox{{$\scriptstyle \sim$}}}}
\def\m@th{\mathsurround=0pt }
\def\eqalign#1{\null\,\vcenter{\openup1\jot \m@th
 \ialign{\strut\hfil$\displaystyle{##}$&$\displaystyle{{}##}$\hfil
 \crcr#1\crcr}}\,}

\title[Wider protocluster environment unveiled]
{ALMA unveils wider environment of distant red protocluster core}

\author[Ivison et al.]{%
\Large 
R.\,J.~Ivison,\(^{\! 1}\) 
A.\,D.~Biggs,\(^{\! 1}\)
M.~Bremer,\(^{\! 2}\) 
V.~Arumugam\(^{3,1}\)
and
L.~Dunne\(^{4}\)
\vspace*{1mm}\\
\(^1\) European Southern Observatory, Karl-Schwarzschild-Strasse~2, D-85748 Garching, Germany\\
\(^2\) School of Physics, University of Bristol, Tyndall Avenue, Bristol BS8~1TL\\
\(^3\) Institut de Radioastronomie Millim\'etrique, 300 rue de la Piscine, F-38406 Saint-Martin d'H\`eres, France\\
\(^4\) School of Physics \& Astronomy, Cardiff University,
Queen's Buildings, The Parade, Cardiff CF24~3AA
}

\date{
Accepted 2020 June 17. Received 2020 June 11; in original form 2020
January 30}

\pubyear{2020}

\begin{document}
\label{firstpage}
\pagerange{\pageref{firstpage}--\pageref{lastpage}}
\maketitle


\begin{abstract}
  We report observations with the Atacama Large Millimetre Array
  (ALMA) of six submillimetre galaxies (SMGs) within 3\,arcmin of the
  Distant Red Core (DRC) at $z=4.0$, a site of intense cluster-scale
  star formation, first reported by \citet{oteo18}.  We find new
  members of DRC in three SMG fields;
  in two fields, the SMGs are shown to lie along the line of
  sight towards DRC; one SMG is spurious.  Although at first sight
  this rate of association is consistent with earlier predictions,
  associations with the bright SMGs are rarer than expected, which
  suggests caution when interpreting continuum over-densities.  We
  consider the implications of all 14 confirmed DRC components passing
  simultaneously through an active phase of star formation.  In the
  simplest explanation, we see only the tip of the iceberg in terms of
  star formation and gas available for future star formation,
  consistent with our remarkable finding that the majority of newly
  confirmed DRC galaxies are not the brightest continuum emitters in
  their immediate vicinity.  Thus while ALMA continuum follow-up of
  SMGs identifies the brightest continuum emitters in each field, it
  does not necessarily reveal all the gas-rich galaxies. To hunt
  effectively for protocluster members requires wide and deep
  spectral-line imaging to uncover any relatively continuum-faint
  galaxies that are rich in atomic or molecular gas. Searching with
  short-baseline arrays or single-dish facilities, the true scale of
  the underlying gas reservoirs may be revealed.
\end{abstract}




\section{Introduction}
\label{introduction}

The galaxies that populate the cores of high-redshift protoclusters
--- where the gas is densest and the likelihood of interactions or
mergers is highest --- are expected to be sites of exceptionally
vigorous star formation \citep[e.g.][]{chiang13, chiang17, muldrew15}.
On their way to becoming the massive, passive, early-type ellipticals
and lenticulars that dominate the red sequence population in
present-day massive clusters, their stellar populations appear to be
built over a comparatively short timescale \citep[$<1$\,Gyr,
e.g.][]{thomas10}.  At times, this star formation may be relatively
unobscured, detectable in the rest-frame ultraviolet (UV), but at its
most intense it is expected to be heavily obscured, emitting mainly in
the rest-frame far-infrared (far-IR).  These developing protocluster
cores should then be most easily identifiable as multiple
submillimetre (submm)-bright dusty star-forming galaxies (DSFGs, also
known as submillimetre galaxies, or SMGs), all at the same redshift,
spanning a megaparsec (Mpc) or more, with a combined star-formation
rate of thousands of M$_\odot$\,yr$^{-1}$.

\citet{negrello17} predict roughly one such protocluster at $z>4$ in
every 100\,deg$^2$, with a combined star-formation rate (SFR) of
$>5,300$\, M$_\odot$\,yr$^{-1}$, though the true expected surface
density is uncertain due to the expected steepness of the IR
luminosity function.  During this highly-active phase, these galaxies
provide key information about the formation and early evolution of
galaxy clusters and their most massive central galaxies
\citep[e.g.][]{ragone-figueroa18}.  At their most extreme, these
structures --- if sufficiently numerous --- may imply dark matter halo
masses that challenge our understanding of the early Universe.

Observational evidence for such structures was both sparse and weak
for a considerable time, based on modest source over-densities
(lacking spectroscopic confirmation) around various AGN/galaxy
populations (e.g.\ radio galaxies and quasars) that were exploited as
likely signposts to dense regions of the early Universe
\citep{ivison00, stevens03, stevens04, debreuck04, priddey08}.  The
number of robust examples --- with unambiguous spectroscopic redshifts
for multiple far-IR-luminous galaxies --- is now growing, however
\citep[e.g.][]{chapman09, hodge13, ivison13, ivison19, dannerbauer14,
  casey15, umehata15, miller18, gg19}, demanding answers to new
mysteries. For example, unless the durations of these colossal
starburst events are considerably longer than conventional wisdom
dictates \citep[][]{greve05, tacconi18}, how can so many supposedly
ephemeral starbursts be seen simultaneously, in the same place?
Significantly longer timescales are possible if the stellar initial
mass function (IMF) is top heavy\footnote{Recent theoretical and
  observational evidence for a top-heavy IMF in starbursts
  \citep{romano17, romano19, zhang18, schneider18, motte18} imply
  significantly fewer stars are formed, but many in the research
  community prefer to think of the IMF as invariant.}, such that SFRs
are lower.  Longer timescales are also inevitable if the reservoir of
gas that fuels star formation is routinely under-estimated, perhaps
because some fraction is dark to us \citep[e.g.][]{balashev17}, and/or
some spends considerable time in the intercluster medium, well outside
the starburst environment, caught in a cycle of outflow and accretion
\citep[e.g.][]{cicone18, peroux19}. In any event, the $\alpha$/[Fe]
ratio of massive, low-redshift early-type cluster galaxies seems
to limit the length of any star-formation period in the progenitor
components of such systems \citep{thomas10}.

The discovery of one of the most extreme known protoclusters in the
early Universe happened almost by accident.  Searching for objects
whose far-IR flux densities rise from 250 to 350\,\um\ and then
onwards to 500\,\um\ --- the so-called `500-\um\ risers'
\citep[e.g.][]{cox11, riechers13, duivenvoorden18} ---
\citet{ivison16} created a sample of `ultrared' galaxies, selected via
$S_{250}, S_{350}$ and $S_{500}$ colour cuts (specifically,
$S_{500}/S_{250} \ge 1.5$ and $S_{500}/S_{350} \ge 0.85$), where
$S_{\lambda}$ is the flux density at $\lambda$\,\um\ as measured by
the Spectral and Photometric Imaging Receiver \citep[SPIRE
--][]{griffin10} on board the {\it Herschel Space Observatory}
\citep{pilbratt10}.  To improve the reliability of the sample and the
associated photometric redshift estimates\footnote{Their colours
  suggest $z \gs 4$, typically.}, which would otherwise have been
compromised more severely by the effects of confusion
\citep{bethermin17}, \citeauthor{ivison16} then imaged these ultrared
galaxies from the ground at 850--870\,\um\ with the Submillimetre
Common-User Bolometer Array 2 \citep[SCUBA-2 --][]{holland13} and/or
the Large APEX Bolometer Camera \citep[LABOCA --][]{siringo09}.
Having thus defined and refined a sample of these `ultrared' galaxies,
the Atacama Large Millimetre Array (ALMA) was used by \citet[][see
also \citealt{fudamoto17}]{oteo18} to determine unambiguous redshifts
via the detection of line emission from $^{12}$CO, H$_2$O, [C\,{\sc
  i}], etc., in spectral scans across the 3-mm atmospheric window
\citep[see also, e.g.,][]{asboth16, riechers17, zavala18}.

After significant work --- involving several line misidentifications,
as described in \S2 and \S3.1 of \citeauthor{oteo18} --- multiple
emission lines\footnote{These included $^{12}$CO(2--1),
  $^{12}$CO(4--3), $^{12}$CO(6--5), H$_2$O($2_{11}$--$2_{02}$) and
  extraordinarily broad (1040\,km\,s$^{-1}$ FWHM) [C\,{\sc i}](1--0).} were ultimately
detected from the reddest galaxy in the \citeauthor{ivison16} sample,
a galaxy so red in $S_{500}/S_{250}$ that it lay off the respective
colour-colour plots: HATLAS\,J004223.5$-$334340 or SGP-354388.

This object, denoted `DRC' hereafter, for Distant Red Core, had
already shown hints of multiplicity, having been spatially resolved by
SCUBA-2 and LABOCA, which have FWHM resolutions of $\gs
13$\,arcsec. Interferometry in band 4 (2\,mm) with ALMA --- designed
to detect $^{12}$CO(6--5) and/or H$_2$O($2_{11}$--$2_{02}$) and
reaching an r.m.s.\ noise level of $\sim 6$\,$\mu$Jy\,beam$^{-1}$ in
continuum --- unveiled 11 components, denoted DRC-1 through DRC-11 by
\citeauthor{oteo18}, with a combined SFR of
6,500\,M$_\odot$\,yr$^{-1}$, mainly concentrated in the brightest
three components. All but one of these 2-mm continuum
detections\footnote{The redshift of DRC-5 remains unknown. It may lie
  in front of DRC, since it has no line detection yet has relatively
  bright continuum emission and is also relatively bright in the
  near-IR. On the other hand, the spectral coverage
  ($\approx1$,000\,km\,s$^{-1}$) is insufficient to rule out cluster
  membership convincingly.} were also seen in one or both of the
targeted lines and are thus members of the cluster, scattered across
$\approx 35$\,arcsec ($\approx 260$\,kpc $\times$ 310\,kpc) in the
plane of the sky --- a clear over-density relative to blank-field
counts \citep[e.g.][]{aravena16, oteo16almacal}. Six of the DRC
components detected at 2\,mm (including DRC-5) are also seen in
continuum at 3\,mm; one is radio-loud, DRC-6, with a flat centimetric
spectrum detected by the Karl G.\ Jansky Very Large Array and the
Australia Telescope Compact Array at 5.5--28\,GHz \citep{oteo18}.

Later ALMA observations, with considerably better spatial resolution
($\approx 0.12$\,arcsec FWHM), further resolved the brightest
component, DRC-1 --- which has a total H$_2$ gas mass of
$\approx 10^{11}$\,M$_\odot$ based\footnote{The $2\times$ correction
  to $M_{{\rm H}_2}$ made by \citet{oteo18} to account for the effect
  of the cosmic microwave background (CMB) on the [C\,{\sc i}](1--0)
  measurements was erroneous. [C\,{\sc i}](1--0) falls on the Wien
  side of the expected spectral line energy distribution such that the
  line--CMB contrast remains high even at $z\sim 4$ \citep[see
  Figure~6 of][and the discussion on pages 12--13] {zhang16}.}  on its
brightness in [C\,{\sc i}](1--0) --- into three fainter galaxies
\citep{oteo17hires}, one of which has amongst the highest reported SFR
densities of any known galaxy,
$\Sigma_{\rm SFR} \approx 2,000$\,M$_\odot$\,yr$^{-1}$\,kpc$^{-2}$
\citep[see also][]{oteo17hires, oteo17almacal2, litke19}.  This
multiplicity may explain the extreme width of the [C\,{\sc i}](1--0)
emission line \citep[see, e.g.,][for another protocluster with a
similarly broad line]{ivison13}, though massive outflows driven by
these intense starbursts (and/or any AGN they may contain) may also
contribute \citep[e.g.,][]{veilleux13, veilleux17, cicone18}.

Imaging at 870\,\um\ of a much wider field, centred on the DRC, was
presented by \citet{lewis18}. These LABOCA data exploited the ultrared
galaxy as a signpost to a potentially dense region of the early
Universe, and were obtained several years before the core was shown to
comprise multiple starburst galaxies. The DRC was one of 22 ultrared
galaxies targeted in this way by \citeauthor{lewis18}, who found an
average over-density of DSFGs of $\ge1.5\times$ [$\ge2\times$] at the
95 [50] per cent confidence level. Towards the DRC,
\citeauthor{lewis18} uncovered a total of nine DSFGs across
124\,arcmin$^2$ (average depth, 1.8\,mJy\,beam$^{-1}$ r.m.s.).  Two
are in the core, i.e.\ those resolved later into 11
galaxies\footnote{We follow \citet{oteo18} in adopting the names A--H
  for the LABOCA sources labelled 1--8 in Figure~13 of
  \citet{lewis18}, respectively.} by ALMA at 2\,mm, and into at least
13 at 870\,\um; six more are within 3\,arcmin of the core, and there
is another one further out: in total, a $2.15^{+0.8}_{-0.5}\times$
over-density, i.e.\ marginally less than half this number would be
expected at this depth in a typical blank field
\citep[e.g.][]{geach17}. Combining their 870-\um\ data with
information at 250, 350 and 500\,\um\ from {\it Herschel}'s SPIRE
instrument, \citeauthor{lewis18} claimed one redshift match with the
core, that being component C; components B and D were predicted to lie
at $z=3.5^{+0.3}_{-0.3}$ and $3.2^{+0.6}_{-0.5}$, respectively (the
sub-components comprising component B were shown by \citealt{oteo18}
to lie at $z=4.002$); E was predicted to lie at $z=1.8^{+0.2}_{-0.2}$,
with $z=2.6^{+0.3}_{-0.3}$ for F. No photometric redshifts were listed
for components G and H; indeed, \citeauthor{lewis18} quoted negative
SPIRE flux densities at the position of H.

Here, we use ALMA to study the dynamics of the regions beyond the core
of this extreme protocluster of galaxies by searching for the
$^{12}$CO(4--3) emission line\footnote{The brighter [C\,{\sc ii}] line
  is not observable towards DRC due to poor atmospheric
  transmission at its observed frequency.} towards the six dusty
starbursts closest to DRC, components C--H, to test whether they
belong to the same structure. We use their line and/or continuum
emission to pinpoint the counterparts in {\it Spitzer} imaging, thus
measuring the total stellar mass, as well as the total SFR and gas
mass.

This paper is organised as follows: \S\ref{sec:observations} describes
the new ALMA observations and our reduction of the
data. \S\ref{sec:results} presents the results, while
\S\ref{sec:discussion} presents our discussion of those results.  We
summarise and draw conclusions in \S\ref{sec:summary}.  Throughout, we
adopt a $\Lambda$-CDM cosmology with $\Omega_{\rmn m} = 0.3$,
$\Omega_\Lambda = 0.7$ and $H_0 = 70$\,km\,s$^{-1}$\,Mpc$^{-1}$, such
that 1\,arcsec corresponds to 7.0\,kpc at $z=4.0$, where the Universe
is 1.5\,Gyr old.

\begin{figure}
\centering 
\includegraphics[width=3.35in]{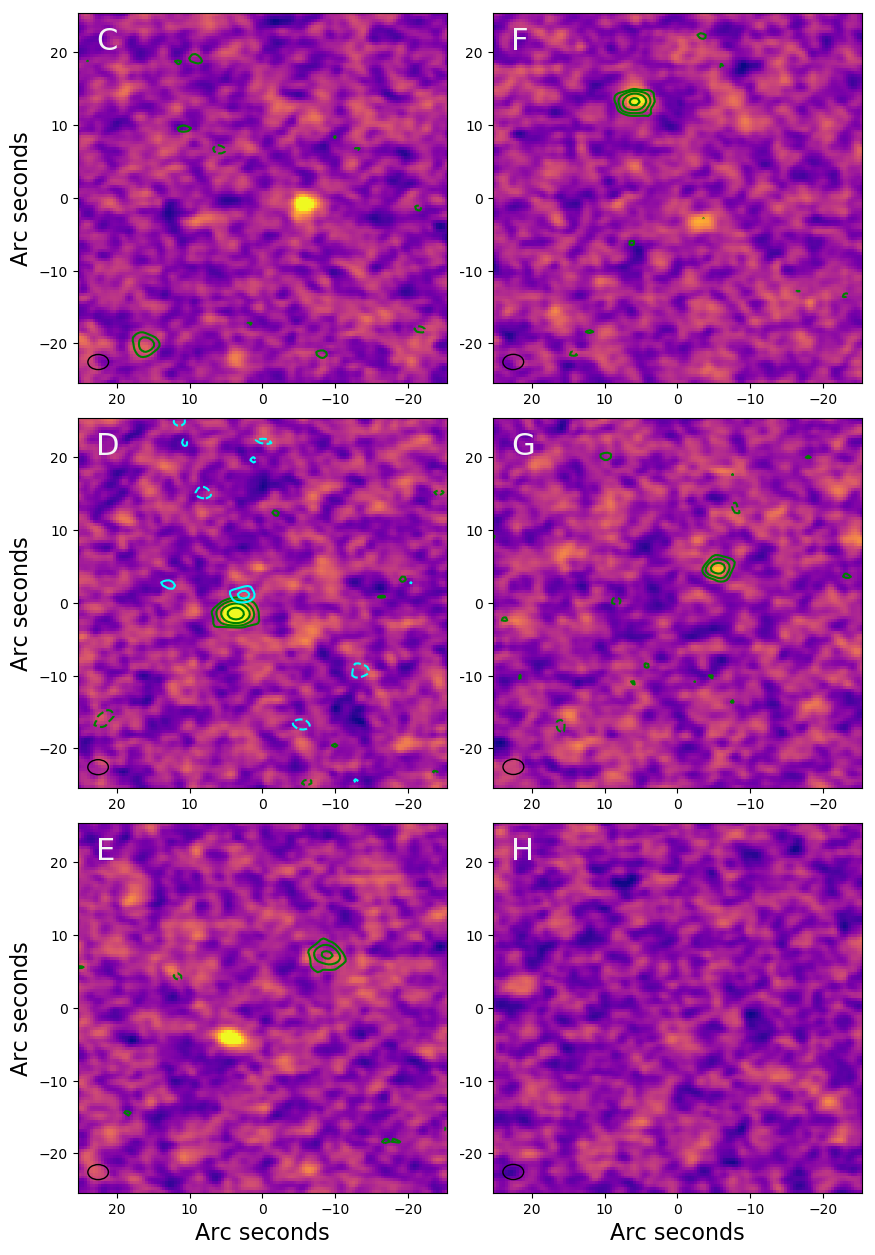}
\caption{ALMA 97-GHz line-free continuum emission with zeroth moment
  spectral-line emission overlaid as contours in green (blue for D)
  for LABOCA sources C--H.  Some or all of the CO(4--3) emission for
  the confirmed members of the protocluster, C, D and E, is {\it not}
  associated with the obvious, bright continuum emitters, except for D
  where the brightest line and continuum emitters are coincident and
  $\approx 2$\,arcsec to the North we see tentative evidence of weak,
  broad line emission (blue contours) offset in velocity by
  $\approx 1$,700\,km\,s$^{-1}$.  The field of source H is devoid of
  both line and continuum emission. The ALMA synthesised beam is shown
  in each panel, lower left. To aid visibility, none of the displayed
  images has been corrected for the primary beam response of the
  antennas.}
\label{fig:linecont}
\end{figure}

\begin{figure}
\centering 
\includegraphics[width=3.35in]{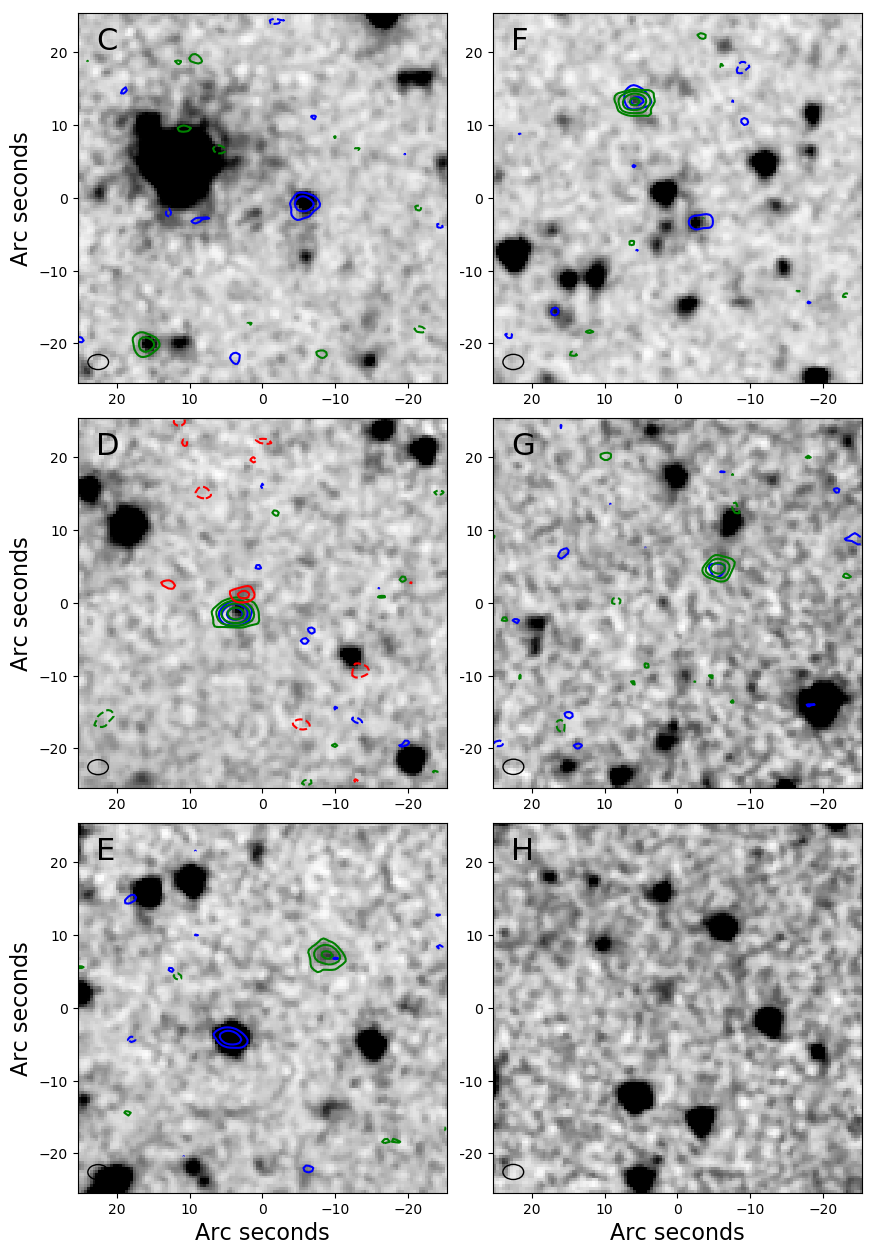}
\caption{IRAC 3.6-$\mu$m images overlaid with ALMA 97-GHz line-free
  continuum contours (blue) and zeroth moment spectral-line contours
  (green, and red for the faint, wide line emission in D).  Except for
  D, the brightest 97-GHz continuum emitters have obvious 3.6-$\mu$m
  counterparts, whereas the CO(4--3) detections associated with weak
  97-GHz continuum emission (C, D, E) have faint emission at
  3.6\,$\mu$m, commensurate with the faint emission seen from the
  other confirmed $z=4$ members of the DRC.  The ALMA synthesised beam
  is shown in each panel, lower left. To aid visibility, none of the
  displayed images has been corrected for the primary beam response of
  the antennas.}
\label{fig:irac}
\end{figure}

\section{ALMA observations and data reduction}
\label{sec:observations}

The six putative cluster members, LABOCA sources C--H, were observed
with ALMA using four dual-polarisation spectral windows in Band~3. One
was used to target $^{12}$CO(4--3) at $z=4.002$ with the remaining
three (and line-free channels from the first) used to measure the
continuum. With 2\,GHz of bandwidth and 128 channels in each spectral
window, we had channel widths of 15.625\,MHz and a velocity resolution
(Hanning-smoothed) of $\sim 100$\,km\,s$^{-1}$ at the observed
frequency of the CO line ($\sim 92.2$\,GHz).

With a primary beam measuring 63\,arcsec (half power beam width), each
DSFG needed to be observed using a separate ALMA pointing centred on
the LABOCA position. The requested sensitivity, $\sigma$, of
0.15\,mJy\,beam$^{-1}$ over a bandwidth of 100\,km\,s$^{-1}$ was
estimated to require about 90\,min of observing time per source and
this was accumulated over 11 separate executions during 2018 April and
July. The 12-m array was moving towards smaller configurations during
this period and the maximum baseline dropped from 629 to 313\,m (the
configurations used were approximately C43-3 and C43-1 in April and
July, respectively).

The data were automatically calibrated using the ALMA Science Pipeline
(part of the Common Astronomy Software Applications package -- CASA)
and each channel was subsequently imaged using the task, {\sc
  tclean}. Natural weighting was used to maximise sensitivity; close
to the expected frequency of the CO line this resulted in a FWHM
synthesised beam of $3.0 \times 2.2$\,arcsec$^2$ at a position angle
(PA) of 91\degr\ (North through East).  Continuum maps were also
created, using channels devoid of line emission, with similar beams to
the line images, with a typical average frequency of 97\,GHz and
r.m.s.\ depths in the range 7.4--8.1\,$\mu$Jy\,beam$^{-1}$. The clean
components found during imaging were used subsequently to create
continuum-free data cubes.

Finally, we searched our data cubes for line emission, starting with the
positions of continuum emission. As it became clear that line emission was
present from sources with no significant 97-GHz continuum, we expanded our
line search to the full cubes. For each source, the final spectrum was
extracted from the spatial pixel with the brightest line emission.

\section{Results}
\label{sec:results}

The ALMA band-3 (97-GHz) line-free continuum and zeroth moment
spectral-line images of our six targets, the LABOCA sources C--H, are
shown in Fig.~\ref{fig:linecont}, and overlaid as contours on
greyscales of deep {\it Spitzer} IRAC 3.6-$\mu$m imaging in
Fig.~\ref{fig:irac}.  The IRAC data are from program IDs: 11107, PI:
P\'erez-Fournon and 13042, PI: Cooray, both of which used 36-position
dither patterns with 30-s exposure times per frame, at slightly
different position angles.  The data were combined with MOPEX using
the standard pipeline.

The basic observational properties of the ALMA band-3 continuum
sources --- positions, peak and total flux densities, where some
continuum sources are clearly resolved spatially by our beam --- are
listed in Table~\ref{tab:contprops}.

The field of LABOCA source H was found to be devoid of continuum
emission and it is therefore likely that the LABOCA detection was
spurious, as already hinted by the empty {\it Herschel} SPIRE maps
reported by \citet{lewis18}.

\begin{figure}
\centering 
\includegraphics[width=3.4in]{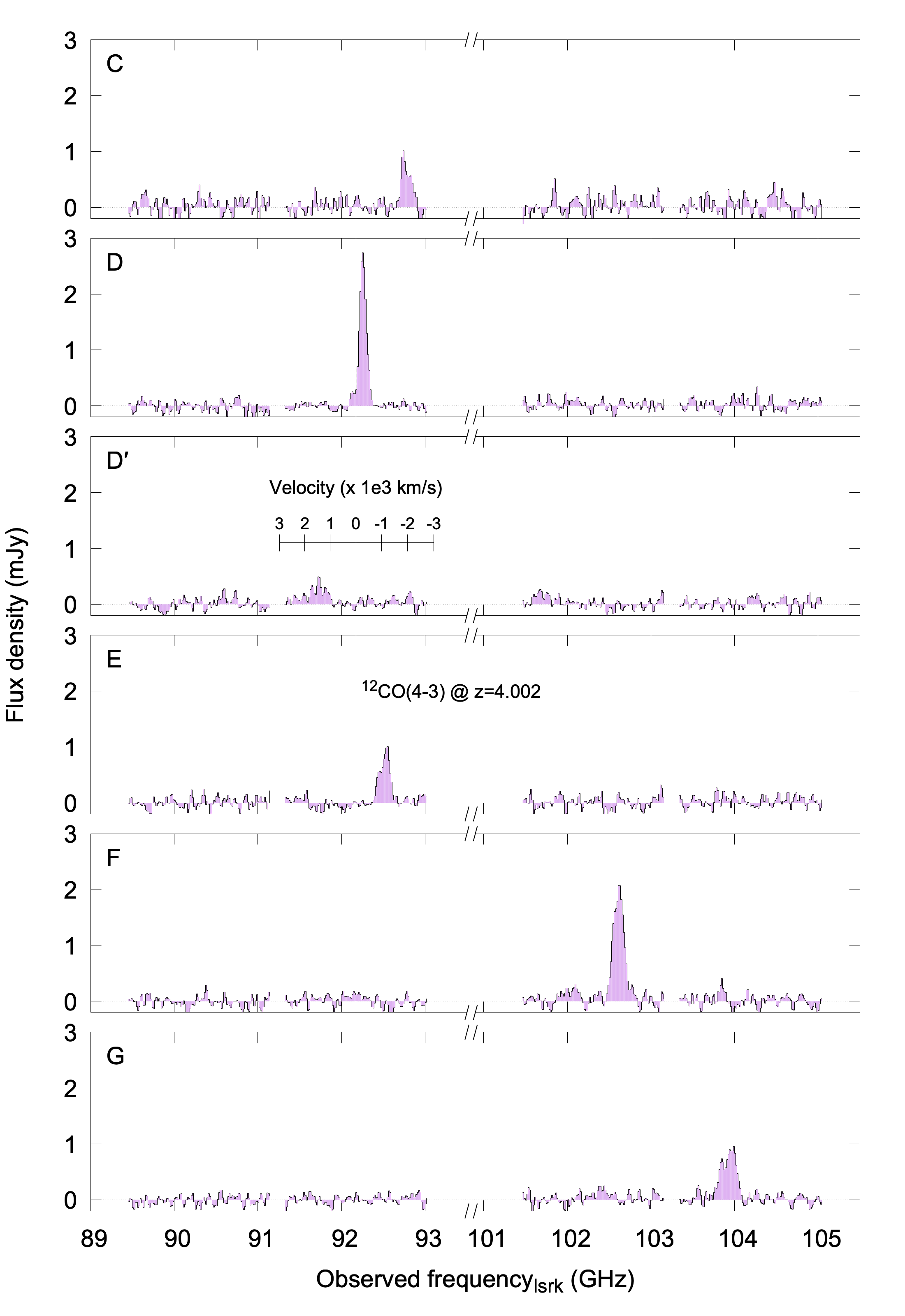}
\caption{ALMA continuum-subtracted spectra of the five line emitters
  found in the fields of
  LABOCA sources C--G.  The field of LABOCA source D was found to
  contain two line emitters, very close together, as described in
  \S\ref{results:d} and as illustrated in Fig.~\ref{fig:linecont},
  with only the brightest line emitter coincident with significant
  continuum emission. Components C, D and E all show $^{12}$CO(4--3)
  emission close to the frequency expected (marked with a vertical
  dotted line) and are thus confirmed as members of the DRC
  protocluster at $z = 4.002$. An additional axis shows velocity
  offsets from the systemic redshift (applicable to all
  sources). Single lines are detected in each of LABOCA sources F and
  G, but the frequency offset is such that these sources cannot be
  associated with the $z=4$ cluster; the identity of each line and its
  associated redshift is unknown. Note that the frequency axis has
  been broken to remove the unobserved 4\,GHz that lies between the
  two ALMA sidebands.}
\label{fig:lines}
\end{figure}

Fig.~\ref{fig:lines} shows our ALMA spectra of the five line emitters
found in the fields of LABOCA sources C--G.  Note that the field of
LABOCA source C was found to contain two faint $>4\sigma$ continuum
sources but the line emitter was found coincident with an even fainter
continuum source, detected barely above the 3-$\sigma$ confidence
level.

For the line emitters in the fields of LABOCA sources C, D and E, the
line frequencies are close to that expected for CO(4--3), marked with
a vertical dotted line in Fig.~\ref{fig:lines}. We regard these as
confirmed members of the DRC protocluster at $z = 4.002$, along with
the ten individual line-emitting components of LABOCA sources A and B
identified in \cite{oteo18}.  The basic observational properties of
these newly-identified lines --- velocity-integrated flux, line width
(FWHM) and the velocity offset from the systemic velocity of the
cluster --- are listed in Table~\ref{tab:lineprops}.

Lines are also detected in each of LABOCA sources F and G, at 102.6
and 103.9\,GHz, respectively, where the frequency offset is such that
these sources cannot be associated with the protocluster; the
identities of these lines and their associated redshifts are currently
unknown.

The resolutions of the ALMA and {\it Spitzer} IRAC data are reasonably
well matched and enable us to pick out unambiguous near-IR
counterparts to the submm/mm continuum emitters trivially.

Except for LABOCA source D, the brightest 97-GHz continuum emitters
have obvious IRAC counterparts, as is usually the case for SMGs
\citep[e.g.][]{egami04}.  On the other hand, the emission-line
detections --- associated with weak 97-GHz continuum emission (sources
C, D, E, F, G) --- show only faint emission at 3.6\,$\mu$m.  This
confirms the reality of these line emitters, if it were ever in doubt,
and hints that F and G also likely lie at higher redshifts than typical
SMGs, $z\gs 3$.

The faintness of the line emitters in the IRAC 3.6 and 4.5\,$\mu$m
filters is commensurate with the other confirmed members of the DRC
protocluster, which are similarly faint.  Of the galaxies discussed by
\citet{oteo18}, only DRC-5 --- the DRC core component without a
line detection and therefore most plausibly a foreground galaxy ---
stands out as being brighter in the near IR.

What is arguably most interesting about these findings is that while
ALMA continuum follow-up of LABOCA sources leads to the identification
of the brightest mm continuum emitters in each field, they do not
necessarily reveal all of the gas-rich galaxies in each field.  If
they are to be effective, it is clear that future searches
for members of protoclusters will need to employ wide-field
spectral-line imaging, targeting lines such as $^{12}$CO, [C\,{\sc i}]
and [C\,{\sc ii}].

\begin{table}
  \begin{center}
    \caption{Properties of the ALMA continuum sources detected
      coincident with the LABOCA sources C--H from \citet{lewis18}. }
  \label{tab:contprops}
\begin{tabular}{lccc} \\ \hline
    Name & $\alpha \;\;\; \;\;\; \;\;\; \;\;\; \;\;\; \;\;\; \delta$ & $S^{\rm tot}_\nu$&Part of\\
               & J2000 &/$\mu$Jy&DRC? \\ \hline
     C$^{\,a}$& 00:42:33.45  $-$33:44:26.4 & $35.8 \pm 11.8$ &Y\\
     D& 00:42:23.55  $-$33:41:18.4 & $151 \pm 9$ &Y\\
     E$^{\,b}$& --- & $3\sigma<50.0$ &Y\\
     F$^{\,c,d}$ & 00:42:20.25 $-$33:44:21.9 & $106 \pm 24^{\,e}$&N\\
     G$^{\,f}$ & 00:42:12.43 $-$33:45:38.9 & $37.8 \pm 9.1$ & N\\
     H & --- & $3\sigma < 30.0$ & N\\ \hline
  \end{tabular}
\end{center}

\noindent
$^a$\,Several brighter continuum sources, neither associated with line
emission (see Figs~\ref{fig:linecont} and \ref{fig:irac}), can be seen
in the field of component C at $\alpha=00$:42:32.70,
$\delta=-$33:44:05.1 J2000 with
$S^{\rm pk}_\nu=75.2 \pm 8.7$\,$\mu$Jy\,beam$^{-1}$ and
$S^{\rm tot}_\nu=98.0 \pm 18.1$\,$\mu$Jy, and at $\alpha=00$:42:33.45,
$\delta=-$33:44:26.4 J2000 with
$S^{\rm tot}_\nu=39.4 \pm 9.8$\,$\mu$Jy.

\noindent
$^b$\,A bright continuum source, unassociated with line emission (see
Figs~\ref{fig:linecont} and \ref{fig:irac}), can be seen in the field
of component E at $\alpha=00$:42:16.46, $\delta=-$33:41:41.9 J2000
with $S^{\rm pk}_\nu=83.0 \pm 8.8$\,$\mu$Jy\,beam$^{-1}$ and
$S^{\rm tot}_\nu=96.8 \pm 16.8$\,$\mu$Jy.  The line
emitter, which lies at $\alpha=00$:42:15.40, $\delta=-$33:41:30.6
J2000, has several faint continuum peaks in its vicinity, none brighter than
27\,$\mu$Jy\,beam$^{-1}$, and an r.m.s.\ noise level of around
9\,$\mu$Jy\,beam$^{-1}$; we ascribe a limit of around 50\,$\mu$Jy at
the $3\sigma$ confidence level.

\noindent
$^c$ Continuum emission with no coincident line emission (see
Figs~\ref{fig:linecont} and \ref{fig:irac}) can be seen in the field
of component F at $\alpha=00$:42:19.53, $\delta=-$33:44:38.5 J2000
with $S^{\rm pk}_\nu=38.5 \pm 7.4$\,$\mu$Jy\,beam$^{-1}$ and
$S^{\rm tot}_\nu=55.1 \pm 16.3$\,$\mu$Jy.

\noindent
$^d$\,Unknown line seen at 102.6\,GHz.

\noindent
$^e$\,$S^{\rm pk}_\nu=58.6 \pm 9.2$\,$\mu$Jy\,beam$^{-1}$. 

\noindent
$^f$\,Unknown line seen at 103.9\,GHz.

\end{table}

\begin{table}
  \begin{center}
    \caption{Properties of Gaussian fits to the three CO(4--3) line
      emitters that are associated with the protocluster, including
      that of the faint, broad companion in field D. Shown are the
      velocity-integrated flux, line width (FWHM), and the velocity
      offset from the systemic velocity of the cluster, taken here to
      be $z=4.002$.}
  \label{tab:lineprops}
  \begin{tabular}{cccc} \\ \hline
    Name & $I_{\rm CO(4-3)}$ & $\Delta\nu$ & FWHM$_{\rm CO(4-3)}$ \\
                        & /mJy\,km\,s$^{-1}$ & /km\,s$^{-1}$ & /km\,s$^{-1}$ \\ \hline
     C & $409 \pm 48$ & $-1929 \pm 16$ & $420 \pm 38$ \\
     D$^a$ & $910 \pm 25$ & $-250 \pm 3$ & $328 \pm 7$ \\
     E  & $456 \pm 31$ & $-1110 \pm 11$ & $480 \pm 25$ \\ \hline
  \end{tabular}
\end{center}

\noindent
$^a$\,The faint emission 3\,arcsec from D measures
$I_{\rm CO(4-3)}=314 \pm 45$\,mJy\,km\,s$^{-1}$,
$\Delta\nu=+1512 \pm 42$\,km\,s$^{-1}$ and
${\rm FWHM}_{\rm CO(4-3)}=923 \pm 100$\,km\,s$^{-1}$.

\end{table}

\subsection{Component D}
\label{results:d}

As well as the five line detections in sources C--E, we tentatively
identify an additional line source, henceforth referred to as
D$^{\prime}$, approximately 3\,arcsec (21\,kpc) North of D, as can be
seen in Figs~\ref{fig:linecont}, \ref{fig:irac} and
\ref{fig:lines}. The peak flux density of the line is
0.5\,mJy\,beam$^{-1}$ ($8\sigma$ in the zeroth moment map); no
associated 97-GHz continuum emission is detected.

Is their small angular separation merely the result of a chance
alignment, or could D$^{\prime}$ and D be physically associated, e.g.\
with D$^{\prime}$ as a fast outflow from D, or with both undergoing
starbursts triggered by a high-velocity fly-by?  The close spatial
proximity suggests a physical association; on the other hand, the
velocity offset between D and D$^{\prime}$, of
$\Delta\nu = 1762 \pm 53$\,km\,s$^{-1}$, where D$^{\prime}$ has a
velocity close to confirmed cluster members C and E, would appear to
rule this out.  The emission is relatively compact and does not extend
back to D, which argues against the outflow hypothesis. Until further
evidence becomes available, we assume that D and D$^{\prime}$ are
separate cluster members, aligned by chance.

\section{Discussion}
\label{sec:discussion}

What have we learnt about the DRC protocluster through the new ALMA
observations described here?

\subsection{Gas mass estimates}
\label{sec:gasmasses}

We estimate the molecular gas mass in the components related to the
DRC using two methods, via the 3-mm continuum and the CO(4--3) line
emission.

The continuum method assumes that the dust is tracing the gas mass ---
presumed to be predominantly molecular in SMGs --- and can either be
calibrated empirically against CO(1--0) for an assumed
$\alpha_{\rm CO}$ \citep[e.g.][]{scoville16} or used directly to
determine a dust mass and hence a gas mass for an assumed gas-to-dust
mass ratio, $\delta_{\rm GDR}$. We convert the observed 3-mm
luminosity to rest-frame 850\,$\mu$m assuming a typical dust SED for
SMGs with $T_{\rm d}=38$\,K and $\beta=1.8$
\citep[e.g.][]{dacunha15}.

For the CO-line method, we calculate the line luminosity and then
assume an excitation, $r_{41}$, and a CO--H$_2$ conversion factor,
$\alpha_{\rm CO}$. The results and assumptions used are listed in
Table~\ref{tab:masses}.

The gas masses inferred from either the 3-mm (97-GHz) dust continuum
or CO(4--3) emission are very large
($\sim10^{11}\,\rm{M_{\odot}}$). The continuum observations are less
sensitive per unit gas mass than the line observations, which means
that in two of the sources (C and E) there is either only a very
marginal detection or an upper limit. With our standard assumptions,
the masses derived from CO(4--3) are higher on average than those
using 3-mm dust continuum (by factors of 2, 1.14 and $>1.8$). This
bias for sources C and E indicates that our standard assumptions are
in some way inappropriate and instead may require a higher $r_{41}$
(0.8), a lower $\alpha_{\rm CO}$ (1.8) or a higher $\delta_{\rm GDR}$
(270), or a combination of these effects.

The cosmic microwave background (CMB) produces a non-negligible
radiation field at $z=4$, where $T_{\rm CMB}=13.7$\,K; this reduces
the observed submm/mm flux densities relative to their intrinsic
values \citep[e.g.][]{dacunha13, zhang16, tunnard17, jin19}.  Observed
97-GHz flux densities correspond to rest-frame 650\,$\mu$m. We do not
have sufficient data to estimate dust temperatures individually for
these sources, but if we assume they are bracketed by the global SED
for the DRC core galaxies ($T_{\rm d}=32$\,K) and the mass-weighted
dust temperature assumed in the calculation of gas mass ($T_{\rm
  d}=25$\,K), the correction to the inferred gas and dust masses would
be 1.3--$1.5\times$ due to the dimming effect of the CMB.

The $J=4$ level of CO is 55\,K above ground with a critical density,
$\sim 1.9 \times 10^4$\,cm$^{-3}$, in the optically thin case. In
lower density regions of the ISM, remote from sites of active star
formation, this line will be sub-thermally excited; any emission we
see in these SMGs is therefore likely to be dominated by the dense and
warm star-forming gas. In the warm and dense gas, the CMB-related
dimming correction to CO(4--3) is negligible, in stark contrast to its
very significant effect on any sub-thermally excited CO(4--3)
emission\footnote{A CMB-induced boost to the CO spectral line energy
  distribution is expected \citep{zhang16}, such that $r_{41}$
  might rise at high redshift. Whether due to an observational
  bias or to the physical effect of the CMB, sources at $z>4$ have
  higher $r_{41}$ than this average: $r_{41}=0.41\pm 0.07$ for SMGs at
  $z=1$--3 \citep{ivison11,bothwell13,cw13}, where $r_{41}=0.76$ for
  G15.141 \citep{cox11}, $r_{41}\sim 1$ for HFLS3 \citep{riechers13},
  $r_{41}=0.45$ for GN20 \citep{carilli10} and $r_{41}=0.43$--0.86 for
  SPT\,2349$-$56 \citep{miller18}.}.

\begin{table}
\begin{center}
  \caption{Luminosities, gas masses and SFRs for the CO(4--3) line
    emitters associated with the protocluster (uncorrected for
    effects of the CMB).}
  \label{tab:masses}
  \begin{tabular}{lccccc} \\ \hline
    Name & $L^{\prime}_{\rm CO(4-3)}$ & $L_{850}$ & $M_{\rm gas}{\rm
                                                    (CO)}$ & $M_{\rm
                                                             gas}{\rm (dust)}$ & SFR \\
                        & /$10^{10}$\,K & /$10^{23}$ & /$10^{11}$\,M$_\odot$ & /$10^{11}$\,M$_\odot$ & /M$_\odot$\\ 
                        & km\,s$^{-1}$ pc$^2$ & W\,Hz$^{-1}$ & & & yr$^{-1}$\\ \hline 
     C & $1.7 \pm 0.2$ & $4.4 \pm 1.5$ & $1.4 \pm 0.2$ & $0.7 \pm 0.2$ & 204\\
     D$^a$ & $3.8 \pm 0.1$ & $18.6 \pm 1.1$ & $3.2 \pm 0.1$ & $2.8 \pm 0.1$ & 485\\
     E  & $1.9 \pm 0.1$ & $<6.2$ & $1.6 \pm 0.1$ & $<0.9$ & 230\\ \hline
  \end{tabular}
\end{center}

\noindent
$^a$\,The faint, broad emission near D has $L^{\prime}_{\rm
  CO(4-3)}=1.29 \pm 0.18$, $L_{850}<3.40$, $M_{\rm gas}{\rm (CO)}=1.10
\pm 0.16$ and $M_{\rm gas}{\rm (dust)}<0.43$.

\noindent
Notes: Observed 3-mm luminosities extrapolated to rest-frame $L_{850}$
using $T_{\rm d}=38$\,K, $\beta=1.8$. Gas masses from CO assume
$r_{41} = 0.41$ \citep{ivison11, bothwell13, cw13} and
$\alpha_{\rm CO}=3.5$. Gas masses from dust assume
$\delta_{\rm GDR}=135$, mass-weighted dust temperature,
$T_{\rm MW}=25$\,K, and $\kappa_{850}=0.065\,\rm{m^2\,kg^{-1}}$. The
luminosities and masses are not corrected for the effects of the CMB
(see \S\ref{sec:gasmasses} for discussion). We suggest a maximum
correction of $1.5\times$ for the dust properties $L_{850}$ and
$M_{\rm gas}{\rm (dust)}$ while we only expect to see CO(4--3) from warm
and dense gas which will not be affected by the CMB dimming.
\end{table}

\subsection{SFRs}
\label{sec:SFRs}

Determining SFRs from far-IR luminosities is problematic in cases
where the 97-GHz flux density is dominated by brighter sources in the
field of the line emitters, such as for LABOCA sources C and E, but we
can instead estimate the SFRs of the line emitter using their CO(4--3)
luminosities, given the strong connection between the luminosity of
the dense gas and SFR \citep{greve14, kennicutt98},
$\rm{log\, SFR = -9.94 +1.08\,log(L^{\prime}_{\rm CO(4-3)})+1.2}$, where
SFR is sensitive to the assumed IMF, as always.

Comparing these SFR estimates to the gas masses inferred independently
from the submm luminosity, we estimate rough, simplistic gas-depletion
timescales for the current starburst events in these systems,
$\approx 60$\,Myr for component D and $\approx 350$\,Myr for C and E.

Overall, the new components of DRC add modestly to its total SFR,
giving a new total, subject to all the usual caveats, of
$\approx 7,400$\,M$_\odot$\,yr$^{-1}$.

\subsection{Protocluster mass}
\label{sec:protoclustermass}

\citet{oteo18} were able to estimate a mass for the protocluster core
from the dynamics of the components identified in that work following
the methods of \citet{evrard08} and \citet{wang16}. While there was
some uncertainty about whether that region would be virialised at
$z=4$, the limited angular (and therefore physical) extent of the
region over which the components with redshifts were identified
($\sim 20$\,arcsec, so a physical radius of $\sim 150$\,kpc) allowed a
plausible estimate of the mass to be determined from the dispersion in
the line-of-sight velocities of the components.  The mass resulting
from consideration of the dynamics was found to be
$\sim 9\times 10^{13}$\,M$_\odot$, roughly $2\times$ higher than the
mass estimate derived for the total IR luminosity from all components
in A and B, and $3\times$ higher than when assuming a simple uniform
spherical distribution of mass encompassing those components and
traced by their velocity dispersion.

Extending these dynamical estimates to include the components C, D and
E, now confirmed to be at $z=4$, is far more uncertain. These
components are of order 100\,arcsec away from the region considered by
\citeauthor{oteo18} --- see their Figure~1. This corresponds to a
projected physical distance of $\sim750$\,kpc, or a comoving distance
$5\times$ larger.  Given the separation and isolation of these
components\footnote{There is no compelling evidence for coherent
  motion across the cluster, just a hint of North-South filamentary
  structure, similar to those seen in the SSA\,22 field by
  \citet{umehata19}, where we also note that the Ly\,$\alpha$ blob
  reported by \citet{oteo18} also runs North--South.}, it seems
unlikely at this redshift that these components are in a virialised
system with those studied by \citeauthor{oteo18}.  Looking at the
velocity offsets relative to the assumed systemic velocity, $z=4.002$,
in Table~2 here and Table~1 of \citeauthor{oteo18}, three of the four
line emitters confirmed in the current work have line-of-sight offsets
of $\pm1,000$--2,000\,km\,s$^{-1}$ while eight of the ten components
with measured redshifts in \citeauthor{oteo18} are within
$\pm500$\,km\,s$^{-1}$. This could be interpreted as an order of
magnitude increase in mass for a virialised system at 100-arcsec
scales relative to that on 20-arcsec scales --- so a mass of several
$10^{14}$\,M$_\odot$, similar or higher than that of the current-day
Virgo cluster, already in place at $z=4$.  This seems unlikely,
however.

It is more likely that these components are embedded in dark-matter
halos that will eventually merge with the (more massive) core traced
by components A and B to form a significantly more massive cluster at
lower redshift. Given their proximity to the core, if we assume an
infall velocity of order 1,000\,km\,s$^{-1}$, and that these halos
have detached from the Hubble flow by $z=4$, they should start to
interact with the core on a timescale of $\approx 1$\,Gyr given the
distance they need to travel. The look-back time to $z=4$ is
11.5\,Gyr, so even if the halos have not yet fully detached from the
universal expansion \citep[unlikely given the simulations reported in,
e.g.,][]{chiang13}, the halos ought to have merged into a larger
system by today, and likely were heading towards virialisation by
$z\sim 1$.

\subsection{On the timing puzzle}
\label{sec:timingpuzzle}

If these newly confirmed components of DRC have yet to interact with
the core, and will not do so for another $\approx 1$\,Gyr, it is
interesting to consider the implications of all 14 DRC components
simultaneously passing through a highly active but comparatively
short-lived phase of star formation.

If there is no causal connection between the activity in each
component, then perhaps the simplest explanation is that we are seeing
the tip of the iceberg in terms of star formation (and gas available
for future star formation) in this system.

If there is significant star-formation activity across this entire
region, at any point in time we are only sensitive to the most extreme
starbursts --- the DSFGs. Given the likely stochastic nature of this
activity, driven in large part by the merging of comparatively massive
gas-rich galaxies that give rise to luminous DSFGs, the galaxies
detectable by observations as sensitive as ours would likely change on
timescales of a few $\times 10$\,Myr.  In the densest regions of DRC,
which we can assume are pinpointed by LABOCA components A and B, there
would always be detectable DSFGs. In less populous sub-halos such as
those marked by components C, D and E, we might see just one or two
DSFGs, or none.  Were we able to carry out an equivalent set of
observations in several 10s of Myr, these components might be
detectable and C, D and E may have vanished. In that case, it is
likely that there are a significant number of other components to this
protocluster that at the time of observation do not have sufficiently
vigorous star-forming galaxies within them to be detected via their
mm/submm continuum emission.

This suggestion is backed up by our remarkable finding that the
majority of newly confirmed members of the DRC protocluster are not
the brightest 3-mm continuum emitters in their respective field, and
are considerably less far-IR luminous than we expected.

We conclude from this that deep, wide-field searches for atomic or
molecular gas in such clusters may uncover a considerable number of
gas-rich but relatively far-IR-faint galaxies.  Indeed, if such
searches are sensitive to emission on $\gs 5$-arcsec scales, via
single-dish facilities or arrays with short baselines, the scale of
the underlying gas reservoir may eventually be revealed.

\section{Summary and concluding remarks}
\label{sec:summary}

We report new ALMA observations of the $z=4$ protocluster known as the
Distant Red Core, or DRC, selected initially by \citet{ivison16} via
its extreme far-IR colours, and reported as a site of intense
cluster-scale star formation by \citet{oteo18}, \citet{lewis18} and
\citet{long20}.

Of the six LABOCA sources reported by \citeauthor{lewis18} within
3\,arcmin of the ten central, confirmed members of the DRC, our ALMA
imaging spectroscopy confirms four new members of the protocluster at
$z=4$ (two towards a single LABOCA source), while two are shown via
robust line detections to lie at other, unknown redshifts along the
line of sight, and one is shown to be a blank submm field.  This rate
of associations is in line with the $\sim2.15\times$ over-density
predicted by \citet{lewis18}, whose individual photometric redshift
estimates are shown to have been useful.

We consider the implications of all 14 confirmed DRC components
passing simultaneously through a highly active but comparatively
short-lived phase of star formation.  Since it is difficult to
envisage a causal connection between the activity in components
separated by 100s of kpc, the simplest explanation is that we are
seeing the tip of the iceberg in terms of star formation (and gas
available for future star formation).  It is likely, then, that other
components of this protocluster are not forming stars with sufficient
vigour at the time of observation to be detected via their mm/submm
continuum emission.  This suggestion is backed up by our remarkable
finding that the majority of newly confirmed members of the DRC
protocluster are not the brightest 3-mm continuum emitters in their
immediate vicinity.  Thus while ALMA continuum follow-up of LABOCA
sources leads to the identification of the brightest mm continuum
emitters in each field, they do not necessarily reveal all of the
gas-rich galaxies. To hunt effectively for members of protoclusters,
we need to employ spectral-line imaging, targeting lines such as
$^{12}$CO, [C\,{\sc i}] and [C\,{\sc ii}]. Searching wide and deep for
atomic or molecular gas in such clusters may uncover a considerable
number of gas-rich but relatively far-IR-faint galaxies. If such
searches are sensitive on $\gs 5$-arcsec scales, via arrays with short
baselines or single-dish facilities, the true scale of the underlying
gas reservoir may eventually be revealed.

\section*{Acknowledgements}

We acknowledge contributions from Alex Lewis and thank Padelis
Papadopoulos for pointing out the erroneous CMB correction in
\citet{oteo18}.

This paper makes use of the following ALMA data:
ADS/JAO.ALMA\#2017.1.00202.S. ALMA is a partnership of ESO
(representing its member states), NSF (USA) and NINS (Japan),
together with NRC (Canada), MOST and ASIAA (Taiwan), and KASI
(Republic of Korea), in cooperation with the Republic of
Chile. The Joint ALMA Observatory is operated by ESO, AUI/NRAO
and NAOJ.

This work is based in part on observations made with the {\it Spitzer
  Space Telescope}, which is operated by the Jet Propulsion
Laboratory, California Institute of Technology under a contract with
NASA.

Funded by the Deutsche Forschungsgemeinschaft (DFG, German Research
Foundation) under Germany's Excellence Strategy --- EXC-2094 ---
390783311.



\bibliographystyle{mnras}
\bibliography{rji}

\bsp

\label{lastpage}

\end{document}